\documentclass[journal]{IEEEtran}

\usepackage{amsmath,graphicx,dsfont}

\usepackage{array}

\usepackage{amsmath}
\usepackage{amssymb}
\usepackage{array}
\usepackage{amsthm}
\usepackage{rotating}
\usepackage{rotfloat}
\usepackage{enumitem}
\usepackage[export]{adjustbox}
\usepackage{tabularx}
\usepackage{multirow}

\usepackage{enumitem}
\usepackage{bm}
\usepackage{textcomp}
\usepackage{subcaption}
\usepackage{graphicx}

\usepackage{booktabs}

\usepackage{tabularx}
\usepackage{amsmath}
\usepackage{tikz}
\usetikzlibrary{shapes,arrows}
\usepackage{verbatim}

\tikzstyle{block} = [draw, rectangle,
minimum height=4em, minimum width=5em]
\tikzstyle{input} = [coordinate]
\tikzstyle{output} = [coordinate]
\tikzstyle{pinstyle} = [pin edge={to-,thin,black}]
\tikzstyle{sum} = [draw, circle, inner sep=3pt]

\begin{document}
\title{Feedback Acquisition and Reconstruction of Spectrum-Sparse Signals by Predictive Level Comparisons}

\author{Mahdi~Boloursaz~Mashhadi,~\IEEEmembership{Student Member,~IEEE,}~Saeed Gazor,~Nazanin Rahnavard,~and~Farokh~Marvasti,~\IEEEmembership{Senior~Members,~IEEE}%
\thanks{Mahdi Boloursaz Mashhadi is both with the ECE Department, University of Central Florida, Orlando, USA and the Advanced Communications Research Institute (ACRI), Sharif University of Technology, Tehran, Iran, e-mail: boloursaz@eecs.ucf.edu.
	
Saeed Gazor is with the ECE Department, Queens' University, Kingston, Canada, Nazanin Rahnavard is with the ECE Department, University of Central Florida, Orlando, USA, and Farokh Marvasti is with the Advanced Communications Research Institute (ACRI), Sharif University of Technology, Tehran, Iran.}}


\maketitle

\begin{abstract}
In this letter, we propose a sparsity promoting feedback acquisition and reconstruction scheme for sensing, encoding and subsequent reconstruction of spectrally sparse signals. In the proposed scheme, the spectral components are estimated utilizing a sparsity-promoting, sliding-window algorithm in a feedback loop. Utilizing the estimated spectral components, a level signal is predicted and sign measurements of the prediction error are acquired. The sparsity promoting algorithm can then estimate the spectral components iteratively from the sign measurements. Unlike many batch-based Compressive Sensing (CS) algorithms, our proposed algorithm gradually estimates and follows slow changes in the sparse components utilizing a sliding-window technique. We also consider the scenario in which possible flipping errors in the sign bits propagate along iterations (due to the feedback loop) during reconstruction. We propose an iterative error correction algorithm to cope with this error propagation phenomenon considering a binary-sparse occurrence model on the error sequence. Simulation results show effective performance of the proposed scheme in comparison with the literature.
\end{abstract}

\begin{IEEEkeywords}
Sparse Signal Acquisition, 1-Bit Compressive Sensing (CS), Level Comparison (LC) Sign Measurements, Binary-Sparse Error Correction.
\end{IEEEkeywords}

\IEEEpeerreviewmaketitle

\section{Introduction}
\IEEEPARstart{S}{pectrum} sparse signals arise in many applications such as cognitive radio networks, frequency hopping communications, radar/sonar imaging systems, musical audio signals and many more. In such cases, the signal components maybe sparsely spread over a wide spectrum and need to be acquired without prior knowledge of their frequencies. This is a major challenge in spectrum sensing that is an essential block in any spectrum-aware communication system. In this research, we propose a scheme and the corresponding signal processing algorithms for acquisition of spectrally sparse signals. The proposed scheme utilizes tools from the general theory of Compressive Sensing (CS)  \cite{Donoho2006CompressedSensing,Candes2008AnSampling} to address spectral sparsity.

Several schemes have already been proposed for sparse signal acquisition. These include the Random Demodulator (RD) \cite{Tropp2010BeyondSignals}, the Multi-coset Sampler \cite{Mishali2009BlindSignals} and the Modulated Wideband Converter (MWC) \cite{Mishali2010FromSignals}. However, the acquired measurements need to be quantized and encoded to bits for subsequent transmission or processing. This is addressed in the Quantized Compressive Sensing \cite{Zymnis2010CompressedMeasurements,Dai2011InformationSensing,Laska2012RegimeSensing} literature.

The extreme case of 1-bit compressive sensing has been extensively studied \cite{Boufounos2009GreedyMeasurements,Laska2011TrustMeasurements,Plan2013RobustApproach,Jacques2013RobustVectors,Li2015RobustErrors,BahmaniRobustPursuit} and proved to be robust against high levels of additive noise on the measurements \cite{Laska2012RegimeSensing}. However, the 1-bit measurements acquired in these works provide no information on the norm of the sparse signal. Hence in these works, reconstruction is possible only up to a scale factor.

In the proposed scheme, the input signal is compared with a level signal \cite{Mashhadi2016IterativeCrossings, marvasti2017wideband, MeIET} and sign measurements of the error are acquired. The level signal is estimated adaptively in a feedback loop utilizing a sparse reconstruction algorithm. The reconstruction algorithm utilizes the previously acquired sign values to estimate the sparse signal components and predict the level signal, subsequently. This overcomes the scale ambiguity of 1-bit CS reconstruction.

The idea of acquiring sign measurements of level comparisons was also applied in \cite{Amini, NormEst, JianLi}. Previous studies on one-bit
sigma-delta quantization \cite{SigmaDelta1,SigmaDelta2,SigmaDelta3} investigate how adaptivity in the level values can improve the reconstruction error bound in terms of the number of measurements. The approach in \cite{SigmaDelta4} achieves exponential decay in the reconstruction error as a function of the number of measurements but requires the levels themselves to be transmitted for reconstruction. This is in contrast to our proposed scheme where the adaptive levels are estimated from the sequence of previously acquired sign measurements themselves. Moreover, unlike many previously proposed batch-based reconstruction algorithms, our proposed algorithm applies one iteration on each sliding window on the input signal using the previous estimate of the sparse vector as an initial estimate. This not only can decrease the computational complexity  for large values of batch sizes and iteration counts, but also enables the proposed algorithm to better follow possible slow changes in the sparse components along iterations. In Section \ref{sec:sim}, we provide performance comparisons with state-of-the-art techniques in \cite{SigmaDelta3, SigmaDelta4} and show effective performance of the proposed scheme by simulations.

In case the acquired sign bits are subsequently transmitted over a channel, the sign bits available to the receiver may contain flipping errors. Due to the feedback, these errors will propagate and make reconstruction unstable. To cope with this, we propose an iterative algorithm for correcting possible sign flip errors assuming a binary-sparse occurrence model on the error sequence. The iterations for error correction are performed along iterations of the main sparse component estimation algorithm at the receiver to gradually estimate the error sequence and avoid error propagation. Unlike the previously proposed error-robust 1-bit CS reconstruction techniques \cite{AOP,Movahed1,Movahed2}, our proposed error correction algorithm alleviates the need for prior knowledge of the number of errors by applying a binary-sparse occurrence model on the error sequence.

This paper is organized as follows. In section \ref{sec:formulation} we describe our proposed feedback acquisition and the corresponding reconstruction scheme. Section \ref{sec:algorithm} presents the algorithms performed in the main building blocks of our proposed scheme. Section \ref{sec:sim} provides the simulation results and finally section \ref{sec:con} concludes the paper.

\section{The Proposed Acquisition and Reconstruction Scheme}
\label{sec:formulation}
In this research, we adopt the sparse exponential model in order to accommodate the general class of spectrally sparse signals that arise in many real world applications. Assuming that power spectrum of $x(t)$ is sparse, we may approximate $x(t)=\sum_{z \in Z} s_z(t)$ as the sum of exponential components for $Z=\{z_1, z_2, \cdots,z_N\}, z_i \in \mathds{C}$ where each component can be predicted by $s_{z_i}(t+\epsilon)=e^{z_i \epsilon}s_{z_i}(t)$. Also assume that $x(t)$ is sparse in the sense that only a few of its components have significant amplitudes $|s_z(t)|$ at any time. Note that the adopted model allows non-equidistant frequencies and hybrid real/imaginary exponentials. 

Fig. \ref{fig:Block diagram} shows the block diagram of the proposed feedback acquisition scheme. In this figure, the complex input signal $x(t)$ is compared with the level signal $\ell(t)$ utilizing a simple comparator. The error signal $e(t)$ goes through the complex sign 
\footnote{The complex sign function is defined as $\mathrm{csgn}(.)=\mathrm{sgn}(\mathrm{Re}(.))+j\mathrm{sgn}(\mathrm{Im}(.))$ where $\mathrm{sgn}(x)=\begin{cases}
1,\qquad &x \ge 0\\
-1,\qquad &x < 0
\end{cases}$, $j=\sqrt{-1}$. $\mathrm{csgn}(.)$ operates element-wisely on vectors.} 
block and is then sampled uniformly at $t=m\tau$ resulting the output sequence of sign values $b_m \in \{\pm1\pm1j\}$. To encode the signal more efficiently, $\ell(t)$ is calculated from $b_m$ in a feedback loop utilizing a sparse component estimation algorithm followed by prediction.

\begin{figure}[h]
	\centering
	\begin{subfigure}[t]{0.45\textwidth}
		\begin{tikzpicture}[auto,node distance=1.8cm,>=latex',scale=.8]
		\node [input, name=input] {};
		\node [sum,right of=input,node distance=1.2cm] (sum) {\Large {+}};
		\node [block,right of=sum,node distance=2.25cm] (csgn) {};
		\draw (csgn)node {\begin{tikzpicture}[scale=1.2]
			\draw[->]
			(0,.1) -- (0,.95);
			\draw (0,1.1) node{csgn};
			\draw[->]
			(-.5,.5) -- (.5,.5);
			\draw[ultra thick,red] (-.5,.25)-- (0,.25) -- (0,.75)--(.5,.75);\end{tikzpicture}};
		\node [block,name=sampler,right of=csgn,node distance=2.75cm,minimum height=0em,minimum width=1em,white,fill=white] {};
		\draw[-] (sampler)+(-.25,0) --+(.25,.5) node[below,near end,pin={[pinstyle]below:{$t=m\tau$}}]{} ;
		\draw [-] (csgn)-- (sampler);
		
		\node[input,name=output,right of=sampler,node distance=1.5cm]{} ;
		\node [block, below of=sampler] (predictor) {\shortstack[c]{Sparse \\ \!\!Component\!\! \\ Estimation}};
		\node [block, below of=csgn] (D/A) {\shortstack[c]{Predict\\ \& \\Hold}};
		\draw [draw,->] (input) -- node[above,near start] {$x(t)$} (sum) node[at end] {+~~~};
		\draw [->] (sum) -- node {$e(t)$} (csgn);
		\draw [->] (sampler) -- (output)--+(.75,0) node[above,midway] {$b_m$};
		\draw [->] (output) |- (predictor);
		\draw [->] (predictor) -- (D/A);
		
		\draw [->] (D/A) -| node[pos=0.99] {$-$}
		node [near end] {$ \ell(t)$} (sum);
	\end{tikzpicture}
	\caption{Block diagram for the proposed acquisition scheme.}
	\label{fig:Block diagram}
	\end{subfigure}
		
	\vspace{.3cm}
	
	\begin{subfigure}[t]{0.45\textwidth}
		\centering
		\begin{tikzpicture}[auto,node distance=1.8cm,>=latex',scale=.8]
		\node [block,right of=sum,node distance=2.25cm] (csgn) {\shortstack[c]{Sparse \\ \!\!Component\!\! \\ Estimation}};
		\node [block,name=sampler,right of=csgn,node distance=2.75cm] {\shortstack[c]{Predict \\ \& \\ Hold}};		
		\draw [->] (csgn)-- (sampler);
		\node [block, below of=csgn] (D/A) {\shortstack[c]{Sparse\\ Error \\Correction}};
		\draw [->] (sum) -- node {$\hat{b}_m$} (csgn);
		\draw [->] (sampler) -- (output)--+(.5,0) node[above,midway] {$\hat{\ell}(t)$};
		\draw [->] (D/A) -- (csgn);
		\draw [->] (csgn) -- (D/A);
		\end{tikzpicture}
		\caption{Block diagram for reconstruction at the receiver.}
		\label{fig:Block diagram2}
	\end{subfigure}
\end{figure}

In many cases, the acquired signal needs to be subsequently transmitted over a channel. In these cases, the sign bits available for reconstruction at the receiver experience flipping errors. These errors cause the receiver to estimate inaccurate level values. If the levels estimated at the receiver are inaccurate, the subsequent sign bits received will be wrongly interpreted which introduces further errors to reconstruction. In other words, due to the feedback, the error propagates and may unstabilize the whole reconstruction. To prevent error propagation, we propose secondary iterations that are applied along iterations of the main sparse component estimation algorithm at the receiver to correct the sign-flip errors as depicted in Fig. \ref{fig:Block diagram2}.

In the next section, we elaborate the algorithms performed in the main building blocks of the proposed scheme.

\section{The Proposed Algorithms}
\label{sec:algorithm}
In this section, we first elaborate our proposed algorithm to be performed in the sparse component estimation block to reconstruct the spectral components from the sign bits. Then, we introduce our proposed sparsity-promoting algorithm to correct sign-flip errors at the receiver. 

\subsection{Sparse Component Estimation}\label{subsec:ComponentEstimation}
Consider a sliding window on the input samples as $X_m=[x(m\tau), x((m-1)\tau),\cdots,x((m-M+1)\tau)]^T$ in which $\tau$ is the sampling period. Moreover, denote the corresponding level and sign values by $L_m=[\ell(m\tau),\ell((m-1)\tau), \cdots\ \ell((m-M+1)\tau)]^T$ and $B_m=[b_m,b_{m-1},\cdots,b_{m-M+1}]^T$, respectively.
Utilizing this vector notation, we get $B_m=\mathrm{csgn}(X_m-L_m)$. Now define $S_m=[s_{z_1}(m\tau),s_{z_2}(m\tau),\cdots,s_{z_N}(m\tau)]^T$ as the state vector for the observed signal $x(t)$, we can write $X_m=\Phi S_{m}$, where $\Phi$ is a Vandermond matrix defined by
\begin{eqnarray}\label{Phi}
\Phi=\left[\begin{array}{cccc}
1&1&\cdots&1\\
e^{-z_1\tau}&e^{-z_2\tau}&\cdots&e^{-z_N\tau}\\
\vdots&\vdots&\ddots&\vdots\\
\!\!\!e^{-z_1(M-1)\tau}\!\!\!&\!\!\!e^{-z_2(M-1)\tau}\!\!\!&\cdots&\!\!\!e^{-z_N(M-1)\tau}\!\!\!
\end{array}\right].
\end{eqnarray}

The exponential modeling $s_{z_i}(t+\epsilon)=e^{z_i \epsilon}s_{z_i}(t)$ simplifies to a   one step predictor as $S_{m}=P \odot S_{m-1}$ where  $P=[e^{z_1\tau}, e^{z_2\tau}, \cdots, e^{z_N\tau}]$ and $\odot$ is element wise multiplication of two vectors.
To estimate and update the sparse state vector $S_{m}$, we propose to iteratively  minimize \begin{align}\label{Lagrange1}
\hat{S}_{m}&=\arg\min_{S}\ \|\hat{B}_{m}-\mathrm{csgn}(\Phi S-L_{m})\|_2^2\\ \nonumber
&+\lambda_1 \|S-P \hat{S}_{m-1}\|_2^2
+\lambda_2 \sum_{i=1}^{N}g_{\sigma}([S]_i),
\end{align}
where $\hat{S}_{m}$ and $\hat{S}_{m-1}$ represent estimates of the vector of sparse components for the sliding windows corresponding to $t=m\tau$ and $t=(m-1)\tau$, respectively, and $[S_{m}]_i$ denotes the $i$th element of the vector $S_{m}$. Note that $\hat{B}_{m}$ is the vector of observed sign bits and is different from the true $B_m$ in the sense that it may contain bit-flip errors. The first term of the cost function in \eqref{Lagrange1} enforces consistency with the encoded sequence of sign values, the second term guarantees smooth update of the solution and the last term promotes sparsity. 

For the sparsity promoting term, we set $g_{\sigma}(s)=\frac{\arctan(\sigma |s|)}{\arctan(\sigma)}$  \cite{Arctan2, Arctan1, Gazor}. It is easy to show that $\lim_{\sigma \to \infty} \sum_ig_{\sigma}([S]_i)=\|S\|_0$ and $\lim_{\sigma \to 0} \sum_ig_{\sigma}([S]_i)=\|S\|_1$. Thus, by starting from a small $\sigma$ value and increasing it along the iterations, we migrate from the convex $\ell_1$ to the non-convex $l_0$ norm gradually. Similarly, for ease of calculating the gradient, we replace the sign function with an S-shaped, infinitely differentiable function \cite{sign1, sign2, zamani}. We set $f_{\delta}(s)=\frac 2{\pi} \arctan(\delta s)$, for some $ \delta>0$ which is differentiable with the derivative  ${f}'(s)=\frac d{ds}f(s)=\frac 2{\pi} \frac{\delta}{1+\delta^2 s^2}$. It is obvious that $\lim_{\delta \to \infty} f_{\delta}(s)=\mathrm{sgn}(s)$ and hence we increase $\delta$ value exponentially along the iterations.
Making these substitutions we get (\ref{Lagrange2}) \footnote{For a function $f:\mathds{R}\mapsto\mathds{R}$, we denote $\mathrm{c}f(.)=f(\mathrm{Re}(.))+jf(\mathrm{Im}(.))$.}

\begin{align}\label{Lagrange2}
\hat{S}_{m}&=\arg\min_{S}\ C(S)\\\nonumber
&=\arg\min_{S} \|\hat{B}_{m}-\mathrm{c}f(\Phi S-L_{m})\|_2^2\\\nonumber
&+\lambda_1 \|S-P \hat{S}_{m-1}\|_2^2
+\lambda_2 \frac{\sum_{i=1}^{N}\arctan(\sigma |[S]_i|)}{\arctan(\sigma)}.\\\nonumber
\end{align}

To solve (\ref{Lagrange2}), we shall find the roots of $\frac{\partial}{\partial S} C(S)=0$. In order to decrease the computational cost, we apply only one iteration on each sliding window but gradually increase the $\sigma$ and $\delta$ parameters along temporal iterations. Utilizing a sliding-window approach also enables following possible changes in the spectral components along iterations. We get,
\begin{align}\label{GradZer1}
&2 \Phi^H {\mathrm{c}f'}(\Phi S-L_m) \odot (\mathrm{c}f(\Phi S-L_m)-\hat{B}_m)\\\nonumber
&+2\lambda_1 (S-P \hat{S}_{m-1})+\frac{\lambda_2}{\arctan(\sigma)}G\odot S=0,
\end{align}
where
\begin{eqnarray}
[G]_i=\frac{1}{|[S]_i|(1+\sigma^2|[S]_i|^2)},\quad \mbox{for } i=1,\cdots,N.
\end{eqnarray}

To solve this non-linear equation, we approximate the first term in (\ref{GradZer1}) by its value at the prior state estimate  and denote $Y_{m-1}=2\lambda_1 P \hat{S}_{m-1}-2 \Phi^H f'(\Phi \hat{S}_{m-1}-L_m) \odot (f(\Phi \hat{S}_{m-1}-L_m)-\hat{B}_m)$, we get (\ref{GradZer2})
\begin{align}\label{GradZer2}
&(2\lambda_1 \mathds{1}_N+\frac{\lambda_2}{\arctan(\sigma)}G)\odot S=Y_{m-1},
\end{align}
where $\mathds{1}_N=[1,\cdots,1]\in\mathds{R}^N$.
The elements of $2\lambda_1 \mathds{1}+\frac{\lambda_2}{\arctan(\sigma)}G$ are $2\lambda_1+\frac{\lambda_2}{\arctan(\sigma)|[S]_i|(1+\sigma^2|[S]_i|^2)}$, which are all real positive values, therefore, from \eqref{GradZer2} we obtain
\begin{align}\label{GradZer3}
\angle[S]_i&=\angle[Y_{m-1}]_i,\\
2\lambda_1|[S]_i|&+\frac{\lambda_2}{\arctan(\sigma)(1+\sigma^2|[S]_i|^2)}=|[Y_{m-1}]_i|.
\label{Eqn: 8}\end{align}

By denoting $\beta=\frac{\lambda_2}{\arctan(\sigma)}$ and $\alpha_i=|[Y_{m-1}]_i|$, we can rewrite \eqref{Eqn: 8} as a cubic polynomial equation in terms of $r_i=|[S]_i|$ given by
\begin{align}\label{GradZer4}
2\lambda_1\sigma^2r_i^3-\alpha_i\sigma^2r_i^2+2\lambda_1r_i+(\beta-\alpha_i)=0.
\end{align}
The coefficients of the cubic polynomial (\ref{GradZer4}) are real. Hence \eqref{GradZer4} has either three real roots or a single real root and a complex conjugate pair. To enforce sparsity, coefficients with smaller amplitudes are encouraged and hence \eqref{Lagrange2} is minimized by choosing the smallest non-negative real root of (\ref{GradZer4}). We propose to solve \eqref{GradZer4} as follows
\begin{description}
  \item[\rm Case 1]  All roots of (\ref{GradZer4}) are real: The sum of the three roots $\frac{\alpha_i\sigma^2}{2\lambda_1\sigma^2}=\frac{\alpha_i}{2\lambda_1}>0$ is always positive and hence there exists at least a positive root. The smallest positive root is feasible for the algorithm.

  \item[\rm Case 2] One of the roots is real and the other two are a complex conjugate pair: If the product of the roots is positive, i.e., $\frac{\alpha_i-\beta}{2\lambda_1\sigma^2}>0$, the real root is  positive and hence feasible. Hence we must enforce $\beta=\frac{\lambda_2}{\arctan(\sigma)}<\alpha_i$ or equivalently increase $\sigma$ such that $\sigma>\arctan(\frac{\lambda_2}{\alpha_i})$. Note that $\sigma$ is already increased along the iterations, hence if this situation happens, $\sigma$ is further increased till $\sigma>\arctan(\frac{\lambda_2}{\alpha_i})$ holds.

\end{description}
As described above, the magnitude and phase of $[\hat{S}_{m}]_i$ are given by the solution of \eqref{GradZer4} and \eqref{GradZer3}, respectively.


Using the state  estimate $\hat{S}_{m}$, the predict \& hold block calculates the next level value as  $\ell((m+1)\tau)=\sum_{i=1}^{N}[P \odot \hat{S}_{m}]_i$. Finally, to get $\ell(t)$ from its samples, each $\ell(m\tau)$ is holded by this block at the output for as long as the sampling period $\tau$.

\subsection{Sparse Error Correction}\label{subsec:ErrPropagation}

Let us define the real and imaginary error vectors $E_m^r$ and $E_m^i$ with elements $e_m^r, e_m^i \in \{0,1\}$. Define $e_m^r=1$ if $\mathrm{Re}(b_m)$ is flipped and $e_m^r=0$, otherwise. Hence, we get $\mathrm{Re}(b_m)=\mathrm{Re}(\hat{b}_m)(1-2e_m^r)$ and $\mathrm{Im}(b_m)=\mathrm{Im}(\hat{b}_m)(1-2e_m^i)$. Note that for ease of calculations, we consider the real and imaginary error vectors separately and provide our algorithm for the real part. The imaginary part is similar. It is obvious that $E_m^r$ itself, is a sparse vector with elements in $\{0,1\}$. Hence, we propose secondary iterations to update and estimate $E_m^r$ along the primary iterations of the sparse component estimation algorithm. Let us denote $\hat{\hat{E}}_{m-1}^r=\mathcal{S}(\hat{E}^r_{m-1})$, in which the $\mathcal{S}(.)$ operator denotes sliding the estimated error vector for one sample and inserting a zero as the initial estimate of its new element. Now to estimate $E^r_{m-1}$, we solve the following
\begin{align}\label{Opt1}
\hat{E}_{m}^r& =\arg\min_{E} h(E) +\theta \sum_{i=1}^{M} [E]_i \\\nonumber
&\textrm{s.t.}\ \quad \|E-\hat{\hat{E}}_{m-1}^r\|_2 \le \epsilon, \quad [E]_i \in [0,1],
\end{align}
where the range for $[E_{m}^r]_i$ is relaxed to be the convex interval $[0,1]$ and the second term of the cost function is the $l_1$ norm which promotes sparsity in $E_{m}^r$ since the elements of $E_{m}^r$  are non-negative. $h(E)=\|\mathrm{Re}(\hat{B}_{m}) \odot (1-2E)-\mathrm{sgn}(\mathrm{Re}(\Phi \hat{S}_{m}-L_{m}))\|_2^2\nonumber $ is a quadratic convex term with regard to $E$.

\begin{table}[h]
	\centering
	\caption{The MSE Values {\rm (dB)} Achieved by the Proposed Scheme}
	\label{MSETable}
	
	\begin{tabular}{r|l|c|c|c|c}
		\multicolumn{2}{c|}{} & $k=2.5$\% &$k=5$\% &$k=10$\%&$k=20$\% \\ \hline
		\multicolumn{2}{c|}{$p=0$} & -19.9 &-19.6 &-17.9&-10.4\\\hline
		\multirow{2}{*}{$p=0.0125$} & w/o EC & -16.3 &-12.1 &-10.3&-6.3\\\cline{2-6}
		& w/ EC & -19.4 & -18.3 &-14.5&-7.8\\ \hline
		\multirow{2}{*}{$p=0.025$} & w/o EC & -10.2 &-8.7 &-4.6&F\\\cline{2-6}
		& w/ EC & -18.2 & -17.4 &-12.5&-7.1\\ \hline
		\multirow{2}{*}{$p=0.05$} & w/o EC & -4.1 &-2.3 &F&F\\\cline{2-6}
		& w/ EC & -17.8 & -16.5 &-10.2&-5.8\\ \hline
	\end{tabular}
\end{table}

To solve (\ref{Opt1}), we use the gradient descent algorithm followed by projection onto $[0,1]$ and stochastic rounding \cite{Raghavan1987RandomizedProofs, 1988ProbabilisticPrograms} to $\{0,1\}$. Note that both the projected gradient and stochastic rounding techniques have convergence guarantees for the convex case as in (\ref{Opt1}). The gradient descent step is given by 
\begin{align}\label{Opt3}
T_{m}=&\hat{\hat{E}}_{m-1}^r-\epsilon \frac{D}{\|D\|_2},
\end{align}
where $\epsilon$ is an small step-size and
\begin{align}
D=&-4 \mathrm{Re}(\hat{B}_{m}) \odot (\mathrm{Re}(\hat{B}_{m)} \odot (1-2\hat{\hat{E}}_{m-1}^r)\\\nonumber
&-\mathrm{sgn}(\Phi \hat{S}_{m}-L_{m}))+\theta \mathds{1}_{M},
\end{align}

The projection and stochastic rounding are performed by
\begin{align}\label{Opt5}
	[E_{m}]_{i}=
	\begin{cases}
		0,   &[T_{m}]_{i} \leq \mathbf{u}\\
		1,   &[T_{m}]_{i} > \mathbf{u},
	\end{cases},
\end{align}
where $\mathbf{u}$ is generated as a uniformly distributed random variable over the interval $[0,1]$.

\section{Simulation results}
\label{sec:sim}
To numerically evaluate the performance of our proposed scheme, we generate random spectrally sparse signals according to the model presented in Section \ref{sec:formulation} with $N=500$, $M=50$, $\tau=5 \times 10^{-4}$ sec, and $Z=\{1j,2j,\cdots,500j\} \times \omega_0$, $\omega_0=10$ rad/sec. The non-zero spectral components are selected uniformly at random and the corresponding amplitudes come from a $\mathcal{N}(0,1)$ distribution. For comparisons, the final normalized reconstruction Mean Square Error ($MSE=10\log_{10}(\frac{\|S-\hat{S}\|_2^2}{\|S\|_2^2})$) values averaged over 100 runs are reported in Table \ref{MSETable} for different sparsity factors. The sparsity factor $k$ is defined as the ratio of the number of nonzero spectral components over the total number of components $N$. The algorithm parameters are experimentally optimized for the best performance as $\delta_m=1.01 \times \delta_{m-1}, \sigma_m=1.1 \times \sigma_{m-1}$.
In this table, $p$ denotes the rate at which sign-flip errors occur, "w/ EC" and "w/o EC" represent the results with and without the proposed error correction (EC) iterations and the letter "F" shows divergence of the proposed algorithm (MSE$>$-5dB) due to the error propagation phenomenon. As shown, EC is necessary to avoid error propagation.

\begin{table}[t]
	\centering
	\caption{MSE Comparisons {\rm (dB)} between our Proposed Scheme and the Literature}
	\label{MSETable1}
	
	\begin{tabular}{c|c|c|c}
		 &$M=50$&$M=100$&$M=200$\\ \hline
		\cite{SigmaDelta3}&-13.34&-18.54&-25.88\\ \hline
		\cite{SigmaDelta4}&-15.76&-29.61&-57.21\\ \hline
		This Work&-16.12&-29.73&-56.96\\ 
	\end{tabular}
\end{table}

Next, we investigate the general scenario in which $x(t)$ both contains frequencies that do not lie on any of the quantized frequencies $Z=\{1j,2j,\cdots,500j\} \times 10$ rad/sec (the off-grid problem) and may also have stable real exponential parts. Note that $exp(\gamma t+j (K\omega_0+\Delta\omega)t)=exp((\gamma+j \Delta\omega) t) \times exp(j K\omega_0 t), \gamma \in \mathds{R}^{-}$ which is the grid frequency $exp(j K\omega_0 t)$ with an amplitude that varies with time according to $exp((\gamma+j \Delta\omega) t)$. Hence, if $\gamma$ and $\Delta \omega$ are small, the algorithm will still be able to converge and follow the smooth changes in the component amplitudes. To investigate this, we generate $x(t)$ with a sparsity factor of $k= 0.05$ that contains components on $\omega={j214.8\times 10, -1.5+j442.1\times 10}$ rad/sec and provide the MSE curves versus iteration in Fig. 2. These curves confirm effective performance of the proposed algorithm to follow smooth changes in the component amplitudes.

\begin{figure}
		\centering
		\includegraphics[scale=.45]{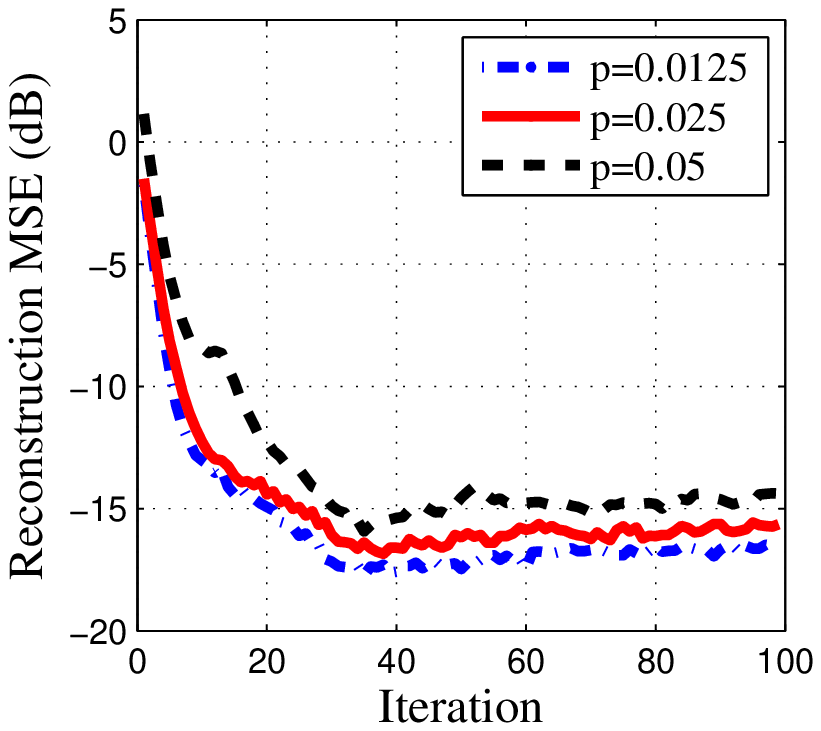}
		\label{MSECurves2}
		\caption{MSE vs. Iteration for the Off-Grid Scenario ($k= 0.05$).}
\end{figure}

Finally in Table \ref{MSETable1}, we compare the performance of our proposed algorithm with state-of-the-art techniques in \cite{SigmaDelta3, SigmaDelta4} for different values of the window length $M$ where $k=5\%$, $p=0$ and the other simulation parameters are fixed as previously. This table provides the final normalized reconstruction MSEs (dB) achieved by the three acquisition/reconstruction schemes averaged over 20 runs when there exists an additive zero-mean Gaussian pre-quantization noise with standard deviation 0.1 and \cite{SigmaDelta4} is applied in a hard thresholding scheme. As observed in this table, both our proposed algorithm and \cite{SigmaDelta4} outperform \cite{SigmaDelta3} especially for larger values of $M$. This is due to an exponential error decay bound for our proposed algorithm and \cite{SigmaDelta4} in comparison with a root exponential decay bound for the $\Sigma\Delta$ scheme in \cite{SigmaDelta3}. Our proposed scheme shows a slightly improved performance in comparison with \cite{SigmaDelta4} for smaller values of $M$ which may be due to improved robustness to noise by the proposed error correction algorithm.

\section{Conclusion}
\label{sec:con}
In this letter, we proposed a feedback acquisition scheme for encoding of spectrally sparse signals to a stream of 1-bit measurements. We proposed a sparsity promoting reconstruction algorithm to predict comparison levels in a feedback loop to facilitate more efficient 1-bit measurements of the input signal. We also proposed a sparse error correction technique to cope with possible sign flip errors during transmission. Finally, we reported simulation results to confirm effective performance of the proposed scheme and algorithms.

\ifCLASSOPTIONcaptionsoff
  \newpage
\fi

\bibliographystyle{IEEEbib}
\bibliography{ref}

\end{document}